\documentstyle[12pt,english,epsf]{article}
\begin{document}
\addtolength{\baselineskip}{-0.3\baselineskip}
\renewcommand{\baselinestretch}{.7}
\noindent
{\large\bf Preparation and measurement: two independent sources
of uncertainty in quantum mechanics}

\noindent
Willem M. de Muynck$^a$\\
{\footnotesize \em Department of Theoretical Physics,
Eindhoven University of Technology, Eindhoven, the Netherlands}\\

\normalsize
\begin{minipage}{13cm}

\noindent
In the Copenhagen interpretation the Heisenberg uncertainty
relation is interpreted as the mathematical expression of the
concept of complementarity, quantifying the mutual disturbance
necessarily taking place in a simultaneous or joint measurement of
incompatible observables. This interpretation has already been
criticized by Ballentine a long time ago, and has recently been challenged
in an experimental way. These criticisms can be substantiated by using the 
generalized formalism of positive operator-valued measures, from
which a new inequality can be derived, precisely
illustrating the Copenhagen concept of complementarity. The
different roles of preparation and measurement in creating
uncertainty in quantum mechanics are discussed.
\end{minipage}

\vspace{.5cm}
\addtolength{\baselineskip}{.3\baselineskip}

\noindent
{\bf I. INTRODUCTION}

The Copenhagen view on the meaning of quantum mechanics largely originated
from the consideration of so-called ``thought experiments'', like the
double-slit experiment and the $\gamma$ microscope. These experiments
demonstrate that there is a mutual disturbance of the measurement
results in a joint measurement of two incompatible 
observables $A$ and $B$ (like position $Q$ and momentum $P$). The
Heisenberg-Kennard-Robertson uncertainty relation 
\begin{equation}
\Delta A\Delta B\geq
\frac{1}{2} \mid \langle [A,B]_-\rangle \mid,
\label{4}
\end{equation}
in which $\Delta A$ and $\Delta B$ are standard deviations,
has often been interpreted as the mathematical expression of
this disturbance (in Heisenberg's paper$^1$ only position $Q$
and momentum $P$ are considered). 
However, as noted by Ballentine$^{2}$, this uncertainty relation
does not seem to have any bearing on the issue of joint
measurement, because it can be experimentally tested by measuring each 
of the observables separately, subsequently multiplying the standard
deviations thus obtained. Moreover, such an interpretation is at
variance with the standard formalism developed by Dirac and von
Neumann, which only allows the 
joint measurement of {\em compatible} observables.
According to Ballentine$^{2}$ quantum mechanics is silent about
the joint measurement of incompatible observables. If this were
true, however, what would this mean for the disturbance idea
originating from the ``thought experiments''? How could these
experiments be useful in clarifying the meaning of a mathematical
formalism that is not capable of yielding a description of such
experiments? 

Nowadays measurements like the double-slit experiment no longer are
``thought'' experiments$^{3-9}$, and complementarity, 
in the sense of mutual disturbance, has been experimentally
demonstrated in an unequivocal way. However, in agreement with
Ballentine's observation the relation of these experiments with
the Heisenberg-Kennard-Robertson inequality (\ref{4}) has
proved controversial$^{10,11}$. Whereas Storey et al.$^{10}$
conclude that ``the principle of complementarity is a
consequence of the Heisenberg uncertainty relation,'' Scully et
al.$^{11}$ observe that ``The principle of complementarity is
manifest although the position-momentum uncertainty relation
plays no role.'' Duerr et al.$^{9}$ stress that quantum
correlations due to the interaction of object and detector,
rather than ``classical'' momentum transfer, enforces the loss
of interference in a which-way measurement. In their
experiment momentum disturbance is not large enough to account
for the loss of interference if the measurement arrangement is
changed so as to yield `which-way' information.

Actually, two different questions are at stake here. First, the question
might be posed whether the Heisenberg inequality of {\em
position and momentum} is the relevant one for interference
experiments. Second, there is the problem observed by
Ballentine, which is the more
fundamental question whether the Heisenberg inequality is
applicable at all. Contrary to the latter question, the former
%xxx Ralston!
might be thought to have a relatively simple answer. In
general, interference experiments like the one of Ref.~$9$ are not joint
measurements of position and momentum but of a different pair
of observables $A$ and $B$ (see section~IV for an example).
Hence, rather than the inequality $\Delta Q \Delta P\geq
\hbar/2$ relation (\ref{4}) for observables $A$ and $B$ seems to
be relevant to the experiment. Although
position and momentum may also be disturbed by the interaction with
the detector, this need not be related to complementarity
because $A$ and $B$ rather than $Q$ and $P$ are involved in the 
correlations between object and detector. Hence, the controversy
could be resolved by pointing out which (incompatible) observables
are measured jointly in the experiment. However,
we would then have to deal with quantum mechanics' alleged silence
with respect to such experiments. 

It seems that Ballentine's problem with respect to the
applicability of (\ref{4}) to the joint measurement of incompatible
observables $A$ and $B$ has more far-reaching consequences because
it points to a fundamental confusion regarding complementarity within
the Copenhagen interpretation. This is due to the poor distinction made
between the different aspects of {\em preparation} and {\em
measurement} involved in physical experimentation. As a matter of fact,
in the Copenhagen interpretation a measurement is not perceived as a
means of obtaining information about the {\em initial}
(pre-measurement) state of the object, but as a way of preparing the
object in some {\em final} (post-measurement) state. Due to this view
on the meaning of ``measurement'' there is insufficient awareness 
that both the preparation of the initial state as well as the
measurement may contribute to the dispersion of an observable. The
Copenhagen issue of complementarity actually has two different aspects,
viz. the aspects of preparation and measurement, which are not
distinguished clearly enough. If such a distinction is duly made, it is
not difficult to realize that the notion of
``measurement disturbance'' should apply to the latter aspect, whereas
the Heisenberg-Kennard-Robertson uncertainty relation refers to the former.
With no proper distinction between preparation
and measurement the Copenhagen interpretation was bound to
amalgamate the two forms of complementarity, thus interpreting the
Heisenberg-Kennard-Robertson uncertainty relation as a property of (joint)
measurement. Unfortunately, remnants of this view are still
abundant in the quantum mechanical literature.

The purpose of the present paper is to demonstrate that the Copenhagen 
confusion of preparation and measurement largely is a consequence of 
the inadequateness of the standard formalism for the purpose of
yielding a description of certain quantum mechanical
experiments, and joint measurements of incompatible
observables in particular. To describe such measurements it is
necessary to generalize the quantum mechanical formalism so as to
encompass positive operator-valued measures$^{12}$ (POVMs);
the standard formalism is restricted to the
projection-valued measures corresponding to the spectral
representations of selfadjoint operators. The generalized
formalism will briefly be discussed in sect~III.
In sect.~IV the generalized formalism will be applied to 
neutron interference
experiments that can be seen as realizations of the double-slit
experiment. By employing the generalized formalism of POVMs it is 
possible to interpret such experiments as joint non-ideal measurements
of incompatible observables like the ones considered in the ``thought
experiments''. An inequality, derived from the generalized theory by
Martens$^{13}$, yields an adequate expression of the mutual 
disturbance of the information obtained on the initial probability
distributions of two incompatible observables in a joint measurement of
these observables. How both contributions to complementarity can be
distinguished in the measurement results obtained in such experiments
will be discussed in sect.~V. A proof of the Martens inequality is 
given in Appendix B. \\

\noindent
{\bf II. CONFUSION OF PREPARATION AND MEASUREMENT}

\noindent
The confusion of preparation and measurement is already present in the
Copenhagen thesis that quantum mechanics is a complete theory.
As a consequence of this thesis a physical quantity cannot have a
well-defined value {\em preceding the measurement} 
(because this would correspond to an ``element of physical reality'' as
employed by Einstein, Podolsky and Rosen$^{14}$ to
demonstrate the {\em in}completeness of quantum mechanics). For this
reason a quantum mechanical measurement cannot serve to ascertain this
value in the way customary in classical mechanics.
Heisenberg$^{15}$ proposed an 
alternative for quantum mechanics, to the effect that the
value of an observable is well-defined {\em immediately after the
measurement}, and, hence, is more or less created by the
measurement$^{16}$. For Heisenberg his uncertainty relation did
not refer to the 
past (i.e. to the initial state), but to the future (i.e. the final
state): it was seen as a consequence of the disturbing influence of the 
measurement on observables that are incompatible with the measured one. 
Hence, for Heisenberg a quantum mechanical measurement was a {\em
preparation} (of the final state of the object), rather than a
determination of certain properties of the initial state. As
emphasized by Ballentine, the interpretation of 
the Heisenberg-Kennard-Robertson uncertainty relation usually found in
quantum mechanics textbooks, 
is in disagreement with Heisenberg's views, because in the textbook
view this relation is not considered a property of the {\em measurement
process} but, rather, of the {\em initial object state}.

Also Bohr$^{17,18}$ did not draw a clear distinction between preparation and 
measurement. He always referred to the {\em complete} experimental
arrangement (often indicated as ``the measuring instrument") serving to
{\em define} the measured observable. For Bohr 
the uncertainty relation (\ref{4}) was an expression of our limitations in
{\em jointly defining} complementary quantities (like
position and momentum) within the context of a measurement. He did not
distinguish different phases of the measurement. More particularly he did
not distinguish different contributions to complementarity 
from the preparation of the initial state and from the disturbance by
the measurement. According to Bohr the uncertainty relation refers to
the ``latitudes'' of the definition of incompatible observables within
the context of a well-defined measurement arrangement, deemed valid for
the measurement as a whole. Incidentally, we see a manifest
difference here with Heisenberg's views, a difference that may have
confused anyone trying to understand 
the Copenhagen interpretation as a consistent way of looking at quantum
mechanics. Moreover, the discrepancy between the Copenhagen
interpretations of the uncertainty relation (viz. as a property of the 
{\em measurement}, either {\em during} this measurement (Bohr), or
afterwards (Heisenberg)) and the textbook interpretation (viz. as a
property of the {\em preparation preceding the measurement}) may have
caused some uneasiness in many students.

Obviously, two completely different issues are at
stake here, corresponding to different forms of complementarity. As
stressed by Ballentine, the Heisenberg-Kennard-Robertson
uncertainty relation (\ref{4}), in 
which $\Delta A$ and $\Delta B$ are standard deviations in separately
performed measurements, should be taken, in agreement with
textbook interpretation, as referring to the {\em preparation of the
initial state}. On the other hand, the Copenhagen idea of complementarity
in the sense of mutual disturbance in a joint 
measurement of incompatible observables, is certainly not without a
physical basis. Thus, in the double-slit experiment (cf. figure~1)
\begin{figure}[t]
\leavevmode
\centerline{
\epsfysize 1.5in
\epsffile{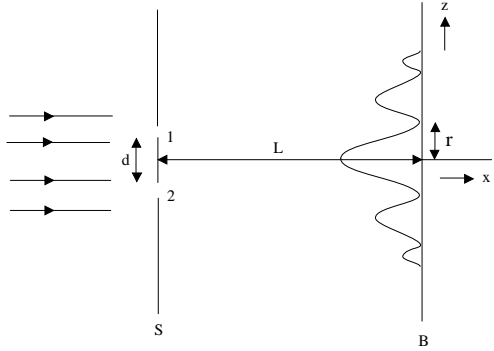}
}
\caption{Double-slit experiment} 
\end{figure}
Bohr demonstrated that, if the quantum mechanical character of
screen $S$ is taken into account, our possibility to define the position 
%xxx Ralston!
and momentum of a particle passing the slits is limited by the
Heisenberg uncertainty relation 
\begin{equation}
\Delta z_{S} \Delta p_{z_S}\geq \hbar/2
\label{5.1}
\end{equation}
of the screen observables $z_{S}$ and $p_{z_S}$. 
As a matter of fact$^{19}$, the lower bounds with which
the latitudes $\delta z$ and $\delta p_z $ of particle position and
momentum are defined, are equal to the standard deviations
$\Delta z_{S}$ and $\Delta p_{z_S}$, respectively. 
Hence, these latitudes must satisfy the inequality
\begin{equation}
\delta z\delta p_z \;\raisebox{-.6ex}{$
{\stackrel{>}{\scriptstyle \sim}}$} \;\hbar/2.
\label{5}
\end{equation}
In Heisenberg's terminology this inequality can be interpreted as
expressing a lower bound for the disturbing influence exerted by the
measuring instrument on the particle, thus causing the {\em
post-}measurement state of the object to satisfy an uncertainty
relation. 

Inequality (\ref{5}) should be distinguished from the 
uncertainty relation
\begin{equation}
\Delta z \Delta p_{z}  \geq \hbar/2
\label{6}
\end{equation}
satisfied by the standard deviations $\Delta z$ and $\Delta
p_{z}$ of position and momentum of the particle in its {\em
initial} state. Whereas inequality (\ref{6}), being an instance
of inequality (\ref{4}), does not refer in any way to joint 
measurement of position and momentum, but can be interpreted as
a property of the {\em preparation of the object preceding the
measurement}, inequality (\ref{5}) does refer to the
{\em measurement process}, since it is derived from a relation
(viz. (\ref{5.1})) satisfied by a part of the measurement
arrangement (screen $S$). 

Unfortunately, in discussions of the double-slit experiment such a
distinction usually is not made. On the contrary, equating the
quantities $\delta z$ and $\delta p_z $ from (\ref{5}) with the
standard deviations $\Delta z$ and $\Delta p_z $, the derivation
of (\ref{5}) is generally interpreted as 
an illustration of the relation (\ref{6}). As a consequence 
it is not sufficiently realized that preparation of the initial state and
(joint) measurement are two distinct physical sources of
uncertainty, yielding similar but physically distinct
uncertainty relations that express different forms of
complementarity. Only the former one is represented by a
relation (viz. (\ref{4})), which can straightforwardly be derived
from the standard formalism. Bohr's analysis of the double-slit
experiment demonstrates that there is a second form of
complementarity, which is {\em not} a property of the
preparation of the initial state as represented by the
Heisenberg-Kennard-Robertson relation, but which is due to
mutual disturbance in a joint measurement of position and momentum. 

One important cause of the mixing up of the two forms of
complementarity is the fact, as stressed by Ballentine, that the
quantum mechanical formalism as axiomatized by von Neumann and
Dirac defies a description of joint measurement
of incompatible observables. In particular, such a measurement would have to
yield joint probability distributions of the incompatible
observables. However, within the standard formalism no
mathematical quantities can be found that are able to play such a role.
Thus, according to Wigner's theorem$^{20}$ no positive phase space
distribution functions $f(q,p)$ exist that are
linear functionals of the density operator $\rho$ such
that $\int dp\; f(q,p) = \langle q |\rho|q \rangle$ and
$\int dq \;f(q,p) = \langle p |\rho|p \rangle$. Also von
Neumann's projection postulate is often interpreted as
prohibiting the joint measurement of incompatible
observables, since there is no unambiguous eigenstate that can
serve as the final state of such a measurement. For this reason
only measurements of one single observable, for which the
Heisenberg-Kennard-Robertson relation has an unambiguous
significance, are usually considered in axiomatic treatments.
%xxx Ralston!

On the other hand, Ballentine's judgment with respect to the
inability of the quantum mechanical formalism to deal with the
second kind of complementarity seems to be too pessimistic. Thus,
for specific measurement procedures {\em generalized} Heisenberg
uncertainty relations have been derived$^{7,8,21,22}$,
different from the Heisenberg-Kennard-Robertson relation, in which the
uncertainties seem to contain contributions from both sources. 
Moreover, in the following it will be demonstrated that 
the generalized quantum mechanical formalism is able to deal with the
two forms of complementarity separately, thus distinguishing 
the contributions due to preparation and (joint) measurement.\\

\noindent
{\bf III. GENERALIZED MEASUREMENTS}

In the generalized quantum mechanical formalism the notion
of a quantum mechanical measurement is generalized so as to encompass
measurement procedures that can be interpreted as
joint measurements of incompatible observables of the type
considered in the ``thought experiments''. A possibility to do
so is offered by the so-called {\em operational
approach}$^{12}$, in which the interaction between object and measuring
instrument is treated quantum-mechanically, and measurement results are
associated with pointer positions of the latter. If $\rho$ and $
\rho_a$ are the initial density operators of object and measuring
instrument, respectively, then the probability of a measurement result
is obtained as the expectation value of the spectral representation
$\{E_m^{(a)}\}$ of some observable of the measuring instrument
in the final state $\rho_f = U \rho \rho_a U^\dagger,\; U=
exp(-iHT/\hbar)$ of the measurement. Thus, $p_m=Tr_{oa} \rho_f 
E_m^{(a)}$. This quantity can be interpreted as a property of the
{\em initial} object state, $p_m=Tr_o \rho M_m$, with $M_m = Tr_{a}
\rho_a U^\dagger E_{m}^{(a)}U $.

The quantum mechanical formalism is generalized to a certain
%xxx Ralston!
extent by the operational approach. Whereas in the standard
formalism quantum mechanical probabilities $p_m$ are represented by
the expectation values of mutually commuting projection
operators ($p_m=\langle E_m\rangle,\;
E_m^2=E_m,\;[E_m,E_{m'}]_-=O $), the generalized formalism
allows these probabilities to be
represented by expectation values of operators $M_m$ that are
not necessarily projection operators, and need not commute
($M_m^2\neq M_m,\; [M_m,M_{m'}]_-\neq O$ in general). The operators 
$M_m,\; O\leq M_m \leq I,\; \sum_m M_m=I$ generate a so-called positive
operator-valued measure (POVM); the observables of the standard
Dirac-von Neumann formalism are restricted to those POVMs of which the
elements are mutually commuting projection operators (so-called
projection-valued measures).

After having generalized the notion of a quantum mechanical observable
it is possible to define a relation of partial ordering between
observables, expressing that the measurement represented by one
POVM can be interpreted as a non-ideal measurement of
another$^{13}$. Thus, we say that a POVM $\{R_m\}$ represents a 
{\em non-ideal} measurement of the (generalized or standard)
observable $\{M_{m'}\}$ if 
the following relation holds between the elements of the POVMs:
\begin{equation}
R_m=\sum_{m'}\lambda_{mm'} M_{m'},\;\lambda_{mm'}
\geq 0,\; \sum_m \lambda_{mm'} = 1.
\label{1}
\end{equation}
The matrix $(\lambda_{mm'} )$ is the non-ideality matrix. It is a
so-called {\em stochastic} matrix$^{23}$.
Its elements $\lambda_{mm'} $ can be interpreted as conditional
probabilities of finding measurement result $a_m$ if an ideal
measurement had yielded measurement result $a_{m'}$.
In the case of an ideal measurement the non-ideality matrix
$(\lambda_{mm'} )$ reduces to the unit 
matrix $(\delta_{mm'})$. As an example we mention photon
counting using an inefficient photon 
detector (quantum efficiency $\eta < 1$), for which the
probability of detecting $m$ photons during a time interval $T$
can be found (cf. Kelley and Kleiner$^{24}$) as:
\begin{equation}
p_m(T) = Tr \rho {\cal N} \left (\frac{(\eta a^\dagger a)^m}{m!}
\exp(-\eta a^\dagger a) \right)
\label{7.2.2}
\end{equation}
(in which $a^\dagger$ and $a$ are photon creation and
annihilation operators, and $\cal N$ is the normal ordering
operator). Defining the POVM $\{R_m\}$ of the inefficient
measurement by means of the equality $p_m(T) = Tr \rho R_m$, it
is not difficult to prove that $R_m$ can be written in the form 
\begin{equation}
R_m = \sum_{n=0}^\infty \lambda_{mn} |n \rangle \langle n|,
\label{7.2.3}
\end{equation}
with $|n \rangle \langle n|$ the projection operator projecting on
number state $|n \rangle$, and 
\begin{equation}
\lambda_{mn} = \left \{
\begin{array}{cl}
0,& m > n,\\
{n\choose m} \eta^m(1-\eta)^{n-m},& m \leq n.
\end{array}\right.
\label{7.2.4}
\end{equation}
For $\eta=1$ the non-ideality matrix is seen to reduce to the
unit matrix, and the POVM (\ref{7.2.3}) to coincide with the
projection-valued measure corresponding to the spectral
representation of the photon number observable 
$N=\sum_{n=0}^\infty n|n \rangle \langle n|$.

Non-ideality relations of the type (\ref{1}) are well-known from
the theory of transmission channels in the classical theory of
stochastic processes$^{25}$, where the non-ideality matrix describes
the crossing of signals between subchannels. 
It should be noted, however, that, notwithstanding the classical origin of
the latter subject, the non-ideality relation (\ref{1}) may be of a
{\em quantum mechanical} nature. Thus, the interaction of the
electromagnetic field with the inefficient detector is a 
quantum mechanical process just like the interaction
with an ideal photon detector is. Relations of the type (\ref{1})
are abundant in the quantum theory of measurement. They can be
employed to characterize the quantum mechanical idea of mutual
disturbance in a joint measurement of incompatible observables. 

Generalizing the notion of quantum mechanical measurement to
the joint measurement of two (generalized) observables, it seems
reasonable to require that such a measurement should yield a
{\em bivariate} joint probability distribution $p_{mn}$,
satisfying $p_{mn}\geq 0, \sum_{mn} p_{mn}=1$. Here $m$ and $n$
label the possible values of the two observables measured
jointly, corresponding to pointer positions of two different pointers
(one for each observable) being jointly read for each individual
preparation of an object. It is assumed that, analogous to the
case of single measurement, the probabilities
$p_{mn}$ of finding the pair $(m,n)$ are represented in the
formalism by the expectation values $\langle R_{mn}\rangle$ of a 
bivariate POVM $\{R_{mn}\},\;R_{mn}\geq O,\;\sum_{mn}R_{mn}=I$
in the initial state of the object.
Then the marginal probabilities $\{\sum_n p_{mn}$ and $\sum_m p_{mn}\}$ 
are expectation values of POVMs $\{M_m=\sum_{n}R_{mn}\}$ and
$\{N_n=\sum_{n}R_{mn}\}$, respectively, which correspond to the
(generalized) observables jointly measured.

In Appendix A it is proven that, if the observables
corresponding to the POVMs $\{M_m\}$ and $\{N_n\}$ are standard
observables (i.e. if the operators $M_m$ and $N_n$ are 
projection operators), then joint measurement is only possible if
these observables commute$^{26}$. This result, derived here from
the generalized formalism, corroborates the standard formalism for
those measurements to which the latter is applicable. Note,
however, that in general commutativity of the operators $M_m$ and $N_n$ is
not a necessary condition for joint measurability of generalized
observables (see section~IV for an example). 

The notion of joint measurement can be extended in the following
way. We say that a measurement, represented by a bivariate POVM
$\{R_{mn}\}$, can be interpreted as a {\em joint non-ideal} measurement of
the observables $\{M_m\}$ and $\{N_n\}$ if the marginals
$\{\sum_nR_{mn}\}$ and $\{\sum_mR_{mn}\}$ of the
bivariate POVM $\{R_{mn}\}$ describing the joint measurement
represent {\em non-ideal} measurements of observables $\{M_m\}$
and $\{N_n\}$. Then, in accordance with (\ref{1}) two non-ideality matrices
$(\lambda_{mm'} )$ and $(\mu_{nn'} )$ should exist, such that
\begin{equation}
\begin{array}{l}
\sum_n R_{mn} = \sum_{m'} \lambda_{mm'} M_{m'},\;\lambda_{mm'}
\geq 0,\; \sum_m \lambda_{mm'} = 1,\\
\sum_m R_{mn} = \sum_{n'} \mu_{nn'} N_{n'},\;\mu_{nn'}\geq
0,\;\sum_{n} \mu_{nn'}=1. 
\end{array}\label{2}
\end{equation}
It is possible that $\{M_m\}$ and $\{N_n\}$ are standard
observables. To demonstrate that the joint measurement
scheme, given above, is a useful one, neutron
interference experiments will be discussed in the next section as
an example, satisfying the definition of a joint non-ideal
measurement of two standard observables. It should be noted that this example 
is not an exceptional one, but can be supplemented by many
others$^{27,28,29,30}$. 
For instance, in analogy ``eight-port optical homodyning$^{5}$''
can be interpreted as a joint non-ideal measurement of the observables $Q=(a +
a^\dagger)/\sqrt{2}$ and $P=(a - a^\dagger)/i\sqrt{2}$ of
a monochromatic mode of the electromagnetic field. 

If $\{M_m\}$ and $\{N_n\}$ are standard observables
the non-idealities expressed by the non-ideality matrices
$(\lambda_{mm'} )$ and $(\mu_{nn'} )$ can be proven$^{13}$ to satisfy
the characteristic traits of the type of complementarity that is
due to mutual disturbance in a joint measurement of incompatible
observables as dealt with in the ``thought experiment''. A measure of
the departure of a non-ideality matrix from the unit matrix is required
for this. A well-known quantity
serving this purpose is Shannon's channel capacity$^{25}$. Here we
consider a closely related quantity, viz. the
average row entropy of the non-ideality matrix $(\lambda_{mm'})$,
\begin{equation}
J_{(\lambda)} = - \frac{1}{N} \sum_{mm'}
\lambda_{mm'} \ln \frac{\lambda_{mm'}} {\sum_{m''}
\lambda_{mm''}},
\label{rowent}
\end{equation}
that (restricting to square $N\times N$ matrices) satisfies the
following properties: 
\[\begin{array}{l}
0 \leq J_{(\lambda)} \leq \ln N,\\
J_{(\lambda)} = 0 \mbox{ \rm if } \lambda_{mm'} = \delta_{mm'},\\
J_{(\lambda)} = \ln N \mbox{ \rm if } \lambda_{mm'} =
\frac{1}{N}. \end{array}\]
Hence, the quantity $J_{(\lambda)} $ vanishes in the case of an
ideal measurement of observable $\{M_{m'}\}$, and obtains its
maximal value if the measurement is uninformative (i.e. does not
yield any information on the observable measured non-ideally) due
to maximal disturbance of the measurement results. 
For a joint non-ideal measurement as defined by (\ref{2}), the
non-idealities of both non-ideality matrices $(\lambda_{mm'} )$
and $(\mu_{nn'})$ can be quantified in a similar way. 

In Appendix B it is demonstrated that for a joint non-ideal
measurement of two standard
observables $A=\sum_m a_m M_m$ and $B=\sum_n b_n N_n$, with
eigenvectors $|a_m \rangle$ and $|b_n \rangle$, respectively,
the non-ideality measures $J_{(\lambda)} $ and $J_{(\mu)}$ obey
the following inequality: 
\begin{equation}
 J_{(\lambda)} + J_{(\mu)} \geq -2\ln \{max_{mn}| \langle a_m|
b_n \rangle|\}.
\label{3} 
\end{equation}
It is evident that (\ref{3}) 
is a nontrivial inequality (the right-hand side unequal to zero)
if the two observables $A$ and $B$ are incompatible in the sense
that the operators do not commute. I shall refer to inequality
(\ref{3}) as the 
{\em Martens inequality}. It is important to note that this inequality
is derived from relation (\ref{2}), and, hence, must be
satisfied in any measurement procedure that can be 
interpreted as a joint non-ideal measurement of two incompatible
standard observables$^{31}$. In relation (\ref{2})
only the {\em observables} (i.e. the measurement procedures) are
involved. Contrary to the Heisenberg-Kennard-Robertson
inequality (\ref{4}), the Martens inequality 
is completely independent of the initial state of the object.
Hence, the Martens inequality does not refer to the preparation of
the initial state, but to the measurement process.

The Martens inequality should be clearly distinguished from 
the entropic uncertainty relation$^{32,33}$ for the standard observables
$A=\sum_m a_m M_m$ and $B=\sum_n b_n N_n$, 
\begin{equation}
H_{\{M_m\}}(\rho) + H_{\{N_n\}}(\rho) \geq -2\ln \{max_{mn}| \langle a_m|
b_n \rangle|\},
\label{10}
\end{equation}
in which $H_{\{M_m\}}(\rho)=-\sum_m p_m\ln p_m,\;p_m=Tr \rho M_m$ (and
analogously for $B$). The inequality (\ref{10}), although
quite similar to the Martens inequality, should be compared with
%xxx Ralston!
the Heisenberg-Kennard-Robertson 
inequality (\ref{4}), expressing a property of the initial
state $\rho$, to be tested by means of separate measurements of
observables $\{M_m\}$ and $\{N_n\}$. \\

\noindent
{\bf IV. NEUTRON INTERFERENCE EXPERIMENTS}

Instead of the classical double-slit experiment we shall consider
an interference experiment performed with
neutrons$^{34,35,36}$. Due to the simplicity of
its mathematical description this experiment yields a
better illustration of the problem of complementarity due to
mutual disturbance in a joint measurement of incompatible
observables, than is provided by the ``thought 
experiment''. The interferometer consists of a silicon crystal with
three parallel slabs (cf. figure~2) in which the neutron
\begin{figure}[t]
\leavevmode
\centerline{
\epsfysize 1.5in
\epsffile{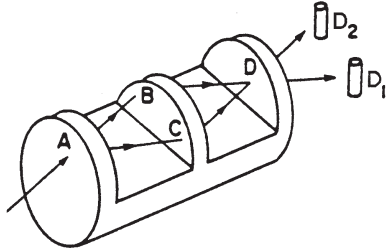}
}
\caption{Neutron interferometer} 
\label{fig7a.1.1}
\end{figure}
can undergo Bragg reflection. A neutron impinging in $A$ at
the Bragg angle is then either transmitted in the same direction or
Bragg reflected. Hence, the neutron may take one of two possible paths.
After reflection in the middle slab ($B$ resp. $C$) the partial waves
of the two paths are brought into interference again in the third slab
($D$). After that the neutron may be found in one of the two out-going
beams by detector $D_1$ or $D_2$. Since it is possible to
achieve a separation of the paths by several centimeters in the
interferometer, it is possible to influence each of the partial beams
separately (cf. figure~3). For instance, we can insert an aluminum
plate into one of the paths, causing a phase shift $\chi$ of the partial
wave, depending on the plate's thickness. By varying the thickness an
interference pattern is obtained when registering the number of
neutrons detected by detector $D_1$ (or $D_2$). 
Summhammer, Rauch and Tuppinger$^{35}$ performed experiments
in which, apart from a phase shifter, an 
absorbing medium was also inserted into one of the paths (indicated in
figure~3 by its transmission coefficient $a$), 
consisting of gold or indium plates. Then the interference pattern
also depends on the value of $a$. The visibility of the interference is
maximal if $a=1$. In such a case we have a pure interference experiment. 
If the absorbing plate is very thick (such that $a=0$) every neutron
taking that path will be absorbed. In that case it is certain that a 
neutron that is registered by one of the detectors has taken the other
path. Then we have a pure ``which path'' measurement, in which the
visibility of the 
interference pattern completely vanishes. For $0<a<1$ the situation is
an intermediate one. This situation will be considered in the
following. Whereas the experiments corresponding to the limiting
values $a=1$ and $a=0$ can be dealt with using the standard
formalism, this is not the case for the
intermediate values of $a$.

Let $|k_1\rangle$ and $|k_2\rangle$ correspond to the plane waves
that impinge at the Bragg angle (cf. figure~3). It is
assumed$^{36}$ that each Bragg reflection induces a phase shift
of $\pi/2$ in the wave vector. The phase shifter changes the phase of
the wave passing it by $\chi$; the absorber alters its amplitude by
a factor of $\sqrt{a}$. Thus for an arbitrary incoming state 
$|in\rangle= \alpha|k_1\rangle+\beta|k_2\rangle,\;
|\alpha|^2+|\beta|^2=1$, we find$^{27}$ the following out-going state:
\begin{equation} 
\begin{array}{lcl}
|out\rangle&=&
\frac{1}{2}[\{\alpha(-1-\sqrt{a}e^{i\chi})+\beta(i-i\sqrt{a}e^{i\chi})\}
|k_1\rangle+\\
&&\{\alpha(-i+i\sqrt{a}e^{i\chi})+\beta(-1-\sqrt{a}e^{i\chi})\}|k_2\rangle]+\\
&&\sqrt{\frac{1-a}{2}}(i\alpha-\beta)|abs\rangle.
\end{array} 
\label{7a.1.9}
\end{equation} 
Here $|abs\rangle$ denotes the state of the absorbed
\begin{figure}[t]
\leavevmode
\centerline{
\epsfysize 1.5in
\epsffile{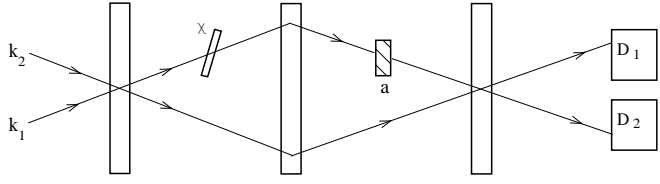}
}
\caption{Neutron interference experiment by Summhammer, Rauch and Tuppinger} 
\label{fig7a.1.3}
\end{figure}
neutron, assumed to be orthogonal to $|k_1\rangle$ and $|k_2\rangle$. 
The detection probabilities $p_1$ and $p_2$ of the two detectors, and the
absorption probability $p_{3}$ are found from this as
\[p_1=|\langle k_1|out\rangle|^2,\;p_2=|\langle
k_2|out\rangle|^2,\; p_{3}=|\langle abs|out\rangle|^2.\]
With  
\begin{equation} 
p_1=\langle in| M_1|in\rangle,\;p_2=\langle in|
M_2|in\rangle,\;p_{3}=\langle in| M_{3}|in\rangle
\label{7a.1.9.1} 
\end{equation} 
the measured detection probabilities are related to the incoming state,
thus yielding an operational definition of the POVM $\{M_1,M_2,M_3\}$
representing the generalized observable measured in the
Summhammer-Rauch-Tuppinger experiment.

We first consider the limits $a=1$ and $a=0$. From (\ref{7a.1.9}) and
(\ref{7a.1.9.1}) for $a=1$ we find that
%xxx Ralston!
$p_3=0$ and $M_1=Q_1,\;M_2=Q_2$, with $Q_1$ and $Q_2$
projection operators, in the two-dimensional representation of the
vectors $|k_1\rangle$ and $|k_2\rangle$ being represented by the matrices
\begin{equation} 
Q_1=\left(\begin{array}{cc}
\cos^2\frac{1}{2}\chi & -\frac{1}{2}\sin\chi\\
-\frac{1}{2}\sin\chi & \sin^2\frac{1}{2}\chi\end{array}\right),\;
Q_2=\left(\begin{array}{cc}
\sin^2\frac{1}{2}\chi & \frac{1}{2}\sin\chi\\
\frac{1}{2}\sin\chi & \cos^2\frac{1}{2}\chi\end{array}\right).
\label{7a.1.6}
\end{equation} 
The standard observable having these operators as its spectral
representation will be referred to as the {\em interference observable}.

For $a=0$ we analogously find $M_1=M_2=P_+/2,\;M_3=P_-$,
with $P_+$ and $P_-$ projection operators represented by the
matrices 
\begin{equation} 
 P_+=\frac{1}{2}\left(\begin{array}{cc}
1 & -i\\
i & 1\end{array}\right),\;
P_-= \frac{1}{2}\left(\begin{array}{cc}
1 & i\\
-i & 1\end{array}\right).
\label{7a.1.8}
\end{equation} 
Also the operators $P_+$ and $P_-$ constitute a spectral representation
of a standard observable, the {\em path observable}, being
incompatible with the interference observable.

For $0<a<1$ the operators $M_m,\;m=1,2,3$ are found in an analogous way,
according to
\begin{eqnarray}
&M_1&=\frac{1}{2}[P_++aP_-+
\sqrt{a}(Q_1-Q_2)],\nonumber\\
&M_2&=\frac{1}{2}[P_++aP_-
-\sqrt{a}(Q_1-Q_2)],\label{7a.1.11}\\
&M_{3}&=(1-a)P_-.\nonumber\end{eqnarray}
It is important to note that in this case the operators $M_1,\;M_2$
and $M_{3}$ are not projection operators. Only in the limits 
$a=1$ and $a=0$ is there a direct link with a standard observable. 
For the majority of experiments ($0<a<1$) the detection
probabilities are not described by the expectation values 
of the spectral representation of one single selfadjoint operator (as
would be the case within the standard formalism). 

It is possible, using definition (\ref{2}), to interpret the
neutron interference experiment
as a joint non-ideal measurement of the interference and path
observables defined by (\ref{7a.1.6}) and (\ref{7a.1.8}).
In order to do so the operators $M_{m}$ of the experiment are 
ordered in a bivariate form according to 
\begin{equation}
 R_{mn}=\left(\begin{array}{cc}
M_1 & M_2\\
\frac{1}{2}M_{3} & \frac{1}{2}M_{3}
\end{array}\right),\;
m=+,-,\;n=1,2.\label{7a.1.13}\end{equation} 
Then the marginals $\{\sum_m
R_{m1},\sum_m R_{m2}\}$ and $\{\sum_n R_{+n},\sum_n R_{-n}\}$ can
easily be verified to satisfy the conditions (\ref{2})
for non-ideal measurements of the path and
interference observables, respectively, with non-ideality matrices 
\begin{equation}
(\lambda_{mm'})=\left(\begin{array}{cc}
1 & a\\
0 & 1-a\end{array}\right),\;
(\mu_{nn'})=\frac{1}{2}\left(\begin{array}{cc}
1+\sqrt{a} & 1-\sqrt{a}\\
1-\sqrt{a} & 1+\sqrt{a}\end{array}\right).\label{7a.1.17}\end{equation}

It is interesting to consider the $a$ dependence of the non-ideality
matrices (\ref{7a.1.17}). For $a=0$ we have
$\lambda_{mm'}=\delta_{mm'},\;\mu_{nn'}=1/2$. In this case
the path measurement is ideal, whereas the non-ideality of the
interference measurement is maximal (the corresponding POVM is given
by $\{I/2,I/2\}$, implying that the POVM's expectation values do not
provide information about the incoming state of the neutron). For $a=1$ the
situation is just the opposite. Then
$\lambda_{+m'}=1, \;\lambda_{-m'}= 0, \;
\mu_{nn'}= \delta_{nn'}$. Now the interference measurement is
ideal, and the path measurement is uninformative. 
For $0<a<1$, in which the standard formalism is not
applicable, both measurements are non-ideal. In going from $a=0$
to $a=1$ the non-ideality of the path measurement increases; that of
the interference measurement decreases.

For the non-ideality measures $J_{(\lambda)}$ and $J_{(\mu)}$
defined by (\ref{rowent}) we obtain
\[\begin{array}{l}
J_{(\lambda)}=\frac{1}{2}[(1+a)\ln(1+a)-a\ln a],\\
J_{(\mu)}=\frac{1}{2}[2\ln2-(1+\sqrt{a})\ln(1+\sqrt{a})-
(1-\sqrt{a})\ln(1-\sqrt{a})]. 
\end{array}\]

\begin{figure}[t]
\leavevmode
\centerline{
\epsfysize 2.2in
\epsffile{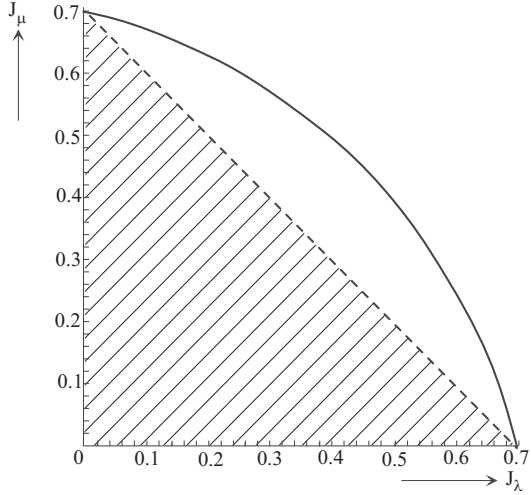}
}
\caption{Parametric plot of $J_{\lambda}$ versus $J_{\mu}$, for $0\leq
a\leq 1$. The shaded area is the region that is forbidden by the
Martens inequality.}
\label{ajp98fig}
\end{figure}
From the parametric plot in figure~4 it can be seen that the Martens
inequality (\ref{3})
is satisfied. This illustrates the impossibility that both
non-ideality measures $J_{(\lambda)}$ and $ J_{(\mu)} $ jointly
have a small value. Figure~4 clearly illustrates the idea of 
complementarity as this arises in the ``thought experiments''. 
If $a$ is varied, then the measurement arrangement is altered. For
Bohr this would signify a different definition of the path and
interference observables for each different value of $a$; the
``latitudes'' of the definition of the observables depending on
$a$. For Heisenberg the path observable is disturbed more by the
measurement process as $a$ increases, whereas the interference 
observable is disturbed less. Both would interpret this as an
expression of the complementarity of the path and interference
observables, due to the fact that the operators $P_+$ and $P_-$ do not
commute with $Q_1$ and $Q_2$. Evidently, the $a$ dependence of
the non-ideality matrices $(\lambda_{mm'})$ and $(\mu_{nn'})$
precisely expresses the complementarity that is connected with
%xxx Ralston!
the mutual disturbance in a joint non-ideal measurement of the
incompatible interference and path observables.

It should be noted, however, that there also is a difference
with Heisenberg's disturbance ideas. In the neutron 
interference experiment the non-idealities do not refer to the final
object state, but to the information obtained on the {\em initial}
state. Hence, these quantities do not refer to the preparative aspect
of measurement (as is the case in Heisenberg's interpretation of his
uncertainty relation), but to the determinative one. Contrary to the
standard formalism the generalized formalism as embodied by
(\ref{2}) is capable
of referring to the past (rather than to the future), even if
measurements are involved in which measurement disturbance plays
%xxx Ralston!
an important role. The generalized formalism enables us to consider
quantum mechanical measurements in the usual determinative
sense, and allows us to distinguish this determinative aspect
from the question in which (post-measurement) state the object
is prepared by the measurement. Incidentally it is noted that
the marginals $\{\sum_m R_{m1},\sum_m R_{m2}\}$ and $\{\sum_n
R_{+n},\sum_n R_{-n}\}$ of the bivariate POVM (\ref{7a.1.13})
constitute a non-commuting pair of generalized observables
jointly measured.\\

\noindent
{\bf V. DISCUSSION}

For unbiased non-ideal measurements, i.e. measurements for which
the non-ideal and the ideal versions in (\ref{1}) yield the same
expectation values for operators $\sum_m a_m M_m$ and $\sum_m a_m R_m$,
the non-ideality matrix 
$(\lambda_{mm'})$ should satisfy the equality $a_{m'}=\sum_m a_m
\lambda_{mm'}$. 
If we restrict to unbiased non-ideal measurements it also is
%xxx Ralston!
possible to demonstrate that there are two sources of
uncertainty by using standard deviations. Thus, using the
notation $r_m=Tr \rho R_{m},\; p_m=Tr \rho 
M_m$, the relation $r_m=\sum_{m'}\lambda_{mm'}
p_{m'}$ between the probability distributions $\{p_m\}$ (of the
ideal measurement) and $\{r_m\}$ (obtained in the non-ideal one)
is found from (\ref{1}). 
For unbiased measurements the standard deviation of the measurement
results $a_m$ of observable $A=\sum_m a_m M_m$, obtained in the non-ideal
measurement, can easily be seen to satisfy the equality
\begin{equation}
\Delta(\{r_m\})^2=\Delta(\{p_m\})^2+\sum_{m'}
\Delta_{m'}^2 p_{m'},
\label{8}
\end{equation}
with
\[
\Delta_{m'}^2:=\sum_m a_m^2 \lambda_{mm'}-(\sum_m a_m
\lambda_{mm'})^2.\] 
The quantity (\ref{8}) consists of two different contributions, i) the
contribution 
$\Delta(\{p_m\})^2$ obtained in an ideal measurement, which is
independent of the parameters of the measurement arrangement, and, for
this reason interpretable in the usual way as a property of the
initial state of the object, and ii) a contribution
$\sum_{m'}\Delta_{m'}^2 p_{m'}$ due to the non-ideality of the
measurement procedure. Also it is not difficult to see that
\[\Delta(\{r_m\})\geq\Delta (\{p_m\}).\]
If in a joint non-ideal measurement of two incompatible
observables $A$ and $B$ both non-ideal measurements are unbiased,
then for the joint non-ideal measurement the {\em 
generalized} Heisenberg uncertainty relation
\begin{equation}
\Delta(\{r_m\})\Delta(\{s_n\})\geq\Delta A\Delta B\geq
\frac{1}{2} \mid \langle [A,B]_-\rangle \mid\label{9}
\end{equation}
($\{s_n\}$ the non-ideally measured probability distribution of
observable $B$) immediately follows from the Heisenberg-Kennard-Robertson
relation (\ref{4}). 

A disadvantage of (\ref{9}) is that not all non-ideal
measurements are unbiased. For instance, as easily follows from
(\ref{7.2.4}), detector inefficiency will cause the average
measured photon number to be smaller than the ideal one. 
For this reason (\ref{9}) is not universally valid. Moreover, 
in the expressions for $\Delta(\{r_m\})$ and
$\Delta(\{s_n\})$ the two contributions to uncertainty 
are merged into one single quantity. 
An inequality, analogous to (\ref{9}), that is valid for
biased measurements too might be obtained by combining the entropic
uncertainty relation (\ref{10}) with the Martens inequality
(\ref{3}), thus yielding 
\begin{equation}
(H_{\{M_m\}} + J_{(\lambda)}) + (H_{\{N_n\}} + J_{(\mu)})\geq
-4\ln \{max_{mn}| \langle a_m| b_n \rangle|\}.\label{11}\end{equation}
However, it is evident that it is not very meaningful to do this
because in (\ref{11}) the two different contributions are once
again  merged, thus veiling their different origins. As follows from
(\ref{10}) and (\ref{3}) both sources satisfy their own inequality.
The 
opportunity entropic quantities offer for exhibiting this seems to
be an important advantage of these quantities over the widely used
standard deviations. It has occasionally been noted$^{22}$ that
for specific measurement procedures an uncertainty 
relation for the joint measurement of incompatible observables
can be formulated in terms of standard deviations. It is not at
all clear, however, whether a relation exists, that is
comparable to the Martens 
inequality, and valid for {\em all} quantum mechanical measurements
interpretable as joint measurements of incompatible standard observables. 

Failure to distinguish the different contributions to
uncertainty represented by the different terms in (\ref{8}) and
(\ref{11}) is at the basis of the Copenhagen confusion
with respect to the uncertainty relations originating with the
discussion of the double-slit experiment. Because no
clear distinction was drawn between preparation and measurement, these
could not be properly distinguished as different sources of
``uncertainty'', {\em both} contributing in their own way. Since
inequality (\ref{5}) refers to the measurement process rather
than to the preparation of the initial state, it should be
compared to the Martens inequality rather than to the
Heisenberg-Kennard-Robertson one.  
The fact that (\ref{5}) has the same mathematical form as (\ref{6}) is
caused by the more or less accidental circumstance that the
uncertainty induced by the measurement process in the
double-slit example is a consequence of the {\em preparation} uncertainty 
of a part of the measurement arrangement (viz. screen $S$) described
by (\ref{5.1}). However, as demonstrated by the neutron
interference example, the measurement disturbance seems to 
more generally originate from the quantum mechanical character of the
whole interaction process of object and measuring instrument.
Fluctuations of the latter may be a part of this, but need not
always play an essential role in the complementarity issue.

It is important to stress that the 
Martens inequality is obtained from the {\em generalized}
formalism, being capable of describing measurements 
represented by POVMs. The founders of the Copenhagen
interpretation did not dispose of this
formalism. Indeed, in the ``thought experiments'' a
measurement is always thought to be represented by a selfadjoint
operator (i.e. a projection-valued measure). In the example of
neutron interferometry this implies a restriction to the 
extreme values $a=0$ and $a=1$. The restriction to
these extreme values was responsible for the view in which
interference is completely disturbed in a `which-way'
measurement (and vice versa). This, indeed, is confirmed by the 
limiting values of the non-ideality matrices (\ref{7a.1.17}), yielding
an uninformative marginal for path if interference is measured ideally
(and vice versa). In the intermediate region $0<a<1$ information on
both observables is obtained, be it that this information is disturbed
in the way described by the Martens inequality.

From the generalized formalism it is clear that in the neutron
interference experiment complementarity of the interference and
path observables (\ref{7a.1.6}) and (\ref{7a.1.8}) is at stake.
Nevertheless, as is evident from the recent discussion referred to
above$^{9,10,11}$, this effect is still sometimes associated with
the Heisenberg inequality for position and momentum. 
It seems that in this discussion the confusion between complementarity
of preparation and measurement still exists. Of course, since a
measurement may also be a preparation procedure for a post-measurement
state of the object, the Heisenberg inequality
$\Delta Q \Delta P\geq \hbar/2$ (as well as inequality (\ref{4})
for any choice of observables $A$ and $B$) should also hold in the
post-measurement object state. This, however, 
is independent of this procedure being a {\em measurement}. As a matter
of fact, $Q$ and $P$ must satisfy the Heisenberg inequality
in the post-measurement state independently of which observables $A$
and $B$ have been measured jointly. 

Complementarity in the sense of mutual disturbance in a joint
measurement of incompatible observables, as characterized by the
Martens inequality, does not refer to the preparation of the
post-measurement state, but to a restriction with respect to
obtaining information on the {\em initial} object state. Apart from  
this difference, the Martens inequality nevertheless seems to be the
mathematical expression of the Copenhagen concept of complementarity,
viz. mutual disturbance in a joint (or simultaneous) {\em
measurement} of incompatible observables. 
It seems that the physical intuition that was expressed by the ``thought
experiments'' was perfect in this respect. However, confusion had to
arise because of the impossibility of dealing with joint measurements
of incompatible observables using the standard formalism.  
Bohr and Heisenberg were led astray by the availability of the
uncertainty relation (\ref{6}) (or, more generally, (\ref{4}))
following from this latter formalism, unjustifiedly 
thinking that this relation provided a materialization of their
physical intuition. \\
%\bibliography{ajp98}

\noindent
{\bf APPENDIX A}

In this Appendix it is proven that standard
observables $A$ and $B$ can be measured jointly if and only if they
commute. Thus, let $M_m$ and $N_n$ be projection operators of
the spectral representations of $A$ and $B$, and 
$M_m=\sum_n R_{mn},\;N_n=\sum_m R_{mn},\; \{R_{mn}\}$ a POVM.
Then $[M_m,N_n]_-=O$, and $R_{mn}=M_mN_n$.

The proof makes use of a well-known property of positive
operators, stating that if $B$ is a positive
operator and $P$ a projection operator satisfying $B \leq P$,
then $B = PBP$. 

Since $R_{mn} \leq M_m$, if $M_m$ is a projection operator we have
$R_{mn} = M_m R_{mn} M_m.$ Since $R_{mn} \leq N_n$, also
$R_{mn} = N_n R_{mn} N_n$.
Hence, $M_m = \sum_{n} R_{mn} = \sum_{n} N_{n} R_{mn} N_{n}.$
Because of $N_n N_{n'} = \delta_{nn'} N_n$, multiplying this
expression from both sides by $N_n$ yields:
\[
M_m N_n = N_n M_m = N_n R_{mn} N_n.
\]
Hence, $[M_m, N_n ]_- = O$, and $R_{mn} = M_m N_n$.

Conversely, if $[M_m, N_n ]_- = O$, then $\{R_{mn} = M_m N_n \}$
is a POVM satisfying
$\sum_n R_{mn} = M_m, \; \sum_m R_{mn} = N_n.$\\

\noindent
{\bf APPENDIX B}

In this appendix a derivation is given$^{13}$ of the Martens
inequality (\ref{3}). We shall restrict ourselves to {\em
maximal} standard observables for which the operators $M_m$ and $N_n$
are one-dimensional projection operators. From
\[
M_m |a_{m'}\rangle = \delta_{mm'} |a_{m'}\rangle,\;N_n |b_{n'}\rangle =
\delta_{nn'} |b_{n'}\rangle \]
and 
\[\sum_{n} R_{mn} = \sum_{m'} \lambda_{mm'} M_{m'},\;\sum_{m} R_{mn} =
\sum_{n'} \mu_{nn'} N_{n'} \]
it follows that
\[\lambda_{mm'} = \langle a_{m'}|\sum_n R_{mn}|a_{m'}\rangle,\;
\mu_{nn'}=  \langle b_{n'}|\sum_m R_{mn}|b_{n'}\rangle.
\]
It is not difficult to see that
$J_{(\lambda)}$ can be written as
\[
J_{(\lambda)} = \frac{1}{N} \sum_{m} (Tr \sum_{n}
R_{mn}) H_{\{M_m\}} \left( 
\frac{\sum_{n'} R_{mn'}}{Tr \sum_{n''} R_{mn''}} \right),\]
and, analogously,
\[J_{(\mu)} = \frac{1}{N} \sum_{n} (Tr \sum_{m}
R_{mn}) H_{\{N_n\}} \left( 
\frac{\sum_{m'} R_{m'n}}{Tr \sum_{m''} R_{m''n}} \right).\]
In these expressions
the arguments of the functions $H_{\{M_m\}}$ and $H_{\{N_n\}}$ are positive
operators with trace equal to $1$. Therefore it is possible to use
the well-known inequality$^{37}$
\begin{equation}\begin{array}{l}
H_{\{M_m\}} (\sum_n p_n\rho_n) \geq
\sum_n p_n H_{\{M_m\}} (\rho_n),\\
O<\rho_n<I,\; Tr \rho_n=1,\;0\leq p_n \leq 1,\; \sum_n p_n=1\end{array}
\label{2.3.1.8}
\end{equation}
to find a lower bound to $J_{(\lambda)}$ (and analogously for
$J_{(\mu)}$).  
Taking in (\ref{2.3.1.8}):
\[ p_n = \frac{Tr R_{mn}}{Tr \sum_{n'} R_{mn'}},\;\rho_n =
\frac{R_{mn}}{Tr R_{mn}}\] 
we obtain the inequality
\begin{equation}\begin{array}{l}
J_{(\lambda)} = \frac{1}{N} \sum_m (Tr \sum_{n'} R_{mn'}) H_{\{M_m\}} \left(
\sum_n p_n \rho_n \right) \geq \\
\frac{1}{N} \sum_{mn} (Tr R_{mn}) H_{\{M_m\}} \left(
\frac{R_{mn}}{Tr R_{mn}} \right).\end{array}
\label{7.10.2.3}
\end{equation}
Analogously we find
\begin{equation}
J_{(\mu)} \geq 
\frac{1}{N} \sum_{mn} (Tr R_{mn}) H_{\{N_n\}} \left(
\frac{R_{mn}}{Tr R_{mn}} \right).
\label{7.10.2.4}
\end{equation}
From (\ref{7.10.2.3}) and (\ref{7.10.2.4}) it then follows that
\[
J_{(\lambda)} + J_{(\mu)} \geq
\frac{1}{N} \sum_{mn} (Tr R_{mn}) \left(H_{\{M_m\}}  \left(
\frac{R_{mn}}{Tr R_{mn}} \right) + H_{\{N_n\}} \left(
\frac{R_{mn}}{Tr R_{mn}} \right) \right).\]
Since also $R_{mn}/Tr R_{mn}$ is a positive operator with trace
$1$, we can use inequality (\ref{10}), with
 $\sum_{mn} R_{mn} = I, \; Tr I = N$ and $ Tr M_m N_n =
|\langle a_m|b_n \rangle |^2$, to arrive at 
the Martens inequality (\ref{3}).
            \begin{tabbing} \` $\Box$ \end{tabbing}

\noindent
{\bf NOTES AND REFERENCES}\\

\noindent
${}^{a)}$W.M.d.Muynck@phys.tue.nl\\
\noindent
${}^{1}$W. Heisenberg, ``Ueber den anschaulichen Inhalt der
quantentheoretischen Kinematik und Mechanik,'' Zeitschr. f.
Phys. {\bf 43}, 172--198 (1927) (English translation: ``The
physical content of quantum kinematics and mechanics,'' in:
J.A. Wheeler and W.H. Zurek, {\em Quantum theory and measurement}
(Princeton University Press, 1983) pp. 62--84).\\ 
\noindent
${}^{2}$L. E. Ballentine, ``The statistical interpretation of
quantum mechanics,'' Rev. Mod. Phys. {\bf 42}, 358--380 (1970).\\
\noindent
${}^{3}$e.g. G. M\"{o}llenstedt and C. J\"{o}nsson,
``Elektronen-Mehrfachinterferenzen an regelm\"assig hergestellten 
Feinspalten,'' Zeitschr. f. Phys. {\bf 155}, 472--474 (1959).\\
\noindent
${}^{4}$J. H. Shapiro and S. S. Wagner, ``Phase and amplitude
uncertainties in heterodyne detection,'' Journ. Quant. Electr.
{\bf QE 20}, 803--813 (1984).\\ 
\noindent
${}^{5}$N. G. Walker and J. E. Caroll, ``Simultaneous phase and
amplitude measurements on optical signals using a multiport
junction,'' Electr. Lett. {\bf 20}, 981--983 (1984).\\
\noindent
${}^{6}$ H. Rauch, ``Tests of quantum mechanics by neutron
interferometry,'' in: G. Gruber et al. eds, {\em Les fondements 
de la m\'ecanique quantique, 25$^e$ Cours de perfectionnement de
l'Association Vaudoise des Chercheurs en Physique} (1983), pp. 330--350; 
G. Badurek, H. Rauch and D. Tuppinger, ``Neutron-interferometric
double-resonance experiment,'' Phys. Rev. {\bf A34}, 2600--2608 (1986).\\
\noindent
${}^{7}$E. Arthurs and J. L. Kelly Jr., ``On the simultaneous
measurement of a pair of conjugate observables,'' Bell Syst.
Techn. Journ. {\bf 44}, 725--729 (1965). \\
\noindent
${}^{8}$Y. Yamamoto, S. Machida, S. Saito, N. Imoto, T.
Yanagawa, M. Kitagawa, and G. Bj\"{o}rk, ``Quantum mechanical
limit in optical precision measurement and communication,'' in:
E. Wolf ed., {\em Progress in optics} (Elsevier Science
Publishers B.V., 1990), Vol.~{\bf XXVIII}, pp. 87--179.\\
\noindent
${}^{9}$S. Duerr, T. Nonn and G. Rempe, ``Origin of
quantum-mechanical complementarity probed by 
a `which-way' experiment in an atom interferometer,''
Nature {\bf 395}, 33-37 (1998).\\
\noindent
${}^{10}$E.P. Storey, S.M. Tan, M.J. Collett and D.F. Walls,
 Nature {\bf 375}, 368 (1995).\\  
\noindent
${}^{11}$M.O. Scully, B.-G. Englert and H. Walther, ``Quantum
optical tests of complementarity,'' Nature {\bf 351}, 111-116
(1991); B.-G. Englert, M.O. Scully and H. Walther, Nature {\bf
375}, 367-368 (1995).\\ 
\noindent
${}^{12}$E. B. Davies, {\em Quantum Theory of Open Systems}
(Academic Press, London, 1976); A. S. Holevo, {\em Probabilistic
and Statistical Aspects of Quantum Theory} (North--Holland,
Amsterdam, 1982); G. Ludwig, {\em Foundations of Quantum
Mechanics} (Springer, Berlin, 1983, Vols. I and II); P. Busch,
M. Grabowski and P. J. Lahti, {\em Operational quantum
mechanics} (Springer-Verlag, Berlin, Heidelberg, 1995).\\
\noindent
${}^{13}$H. Martens and W. de Muynck, ``Nonideal quantum measurements,''
Found. of Phys. {\bf 20}, 255--281 (1990); H. Martens and W. de Muynck,
``The inaccuracy principle,'' Found. of Phys. {\bf 20}, 357--380 (1990).\\
\noindent
${}^{14}$A. Einstein, B. Podolsky, and N. Rosen, ``Can
quantum-mechanical description of physical reality be considered
complete?'' Phys. Rev. {\bf 47}, 777--780 (1935).\\
\noindent
${}^{15}$W. Heisenberg, {\em The physical principles of quantum
theory} (Dover Publications, Inc., 1930), pp. 20--46.\\
\noindent
${}^{16}$See also P. Jordan, ``Quantenphysikalische Bemerkungen zur
Biologie und Psychologie,'' Erkenntnis {\bf 4}, 215--252 (1934). It
should be noted that this does certainly not agree with Bohr's views.\\
\noindent
${}^{17}$N. Bohr, Como Lecture, ``The quantum postulate and the recent
development of atomic theory,'' in: J. Kalckar ed., {\em N. Bohr,
Collected Works} (North--Holland, Amsterdam, 1985), Vol.~6, pp.
113--136.\\ 
\noindent
${}^{18}$N. Bohr, ``Discussion with Einstein on epistemological
problems in atomic physics,'' in: P. A. Schilpp ed., {\em
``Albert Einstein: Philosopher--Scientist''} (Open Court, La
Salle, Ill, 1982), 3rd ed., pp. 199-241.\\
\noindent
${}^{19}$See, for instance, E. Scheibe, {\em The logical analysis of
quantum mechanics} (Pergamon Press, Oxford, etc., 1973), pp. 42--49.\\ 
\noindent
${}^{20}$E.P. Wigner, ``Quantum-mechanical distribution functions
revisited,'' in: W. Yourgrau, A. van der Merwe eds., {\em Perspectives
in quantum theory} (The MIT Press, Cambridge, Mass., 1971), pp. 25--36.\\
\noindent
${}^{21}$E. Arthurs and M. S. Goodman, ``Quantum correlations: A
generalized Heisenberg uncertainty relation,'' Phys. Rev. Lett.
{\bf 60}, 2447--2449 (1988).\\ 
${}^{22}$M. G. Raymer, ``Uncertainty principle for joint
measurement of noncommuting variables,'' Am. J. Phys. {\bf 62},
986--993 (1994). \\
\noindent
${}^{23}$e.g. F.R. Gantmacher, {\em Application of the theory of
matrices} (Interscience Publishers Inc., New York, 1959).\\
\noindent
${}^{24}$P.L. Kelley and W.H. Kleiner, ``Theory of
electromagnetic field measurement and photoelectron counting,''
Phys. Rev. {\bf A136}, 316--334 (1964).\\
\noindent
${}^{25}$e.g. R. McEliece, {\em The theory of information and coding}
(Addison--Wesley, London, 1977).\\
\noindent
${}^{26}$A different proof can be found in A. S. Holevo, {\em
Probabilistic and Statistical Aspects of Quantum Theory}
(North--Holland, Amsterdam, 1982), p. 71. Using the method employed in
Appendix A it is possible to prove that commutativity is necessary and
sufficient for joint measurement already if only one of the marginals is a
standard observable (W.M. de Muynck and J.M.V.A. Koelman, ``On
the joint measurement of incompatible observables in quantum
mechanics,'' Phys. Lett. {\bf 98A}, 1--4 (1983)). \\
\noindent
${}^{27}$W. M. de Muynck and H. Martens, ``Neutron
interferometry and the joint measurement of incompatible
observables,'' Phys. Rev. {\bf 42A}, 5079--5086 (1990).\\
\noindent
${}^{28}$W. M. de Muynck, W. W. Stoffels, and H. Martens,
``Joint measurement of interference and path observables in
optics and neutron interferometry,'' Physica {\bf B 175},
127--132 (1991).\\ 
\noindent
${}^{29}$H. Martens and W. M. de Muynck, ``Single and joint spin
measurements with a Stern-Gerlach device,'' Journ. Phys. A:
Math. Gen. {\bf 26}, 2001--2010 (1993).\\
\noindent
${}^{30}$W. M. de Muynck and H. Martens, ``Quantum mechanical
observables and positive operator valued measures,'' in: W.
Florek, D. Lipinski and T. Lulek eds., {\em Symmetry and
structural properties of condensed matter} (World Scientific,
Singapore, 1993), pp. 101--120.\\ 
\noindent
${}^{31}$Note that the inequality can easily be generalized to
nonmaximal standard observables having a discrete
spectrum. From the example of ``eight-port optical homodyning''
it is seen that also in case of continuous spectra a similar
behaviour of mutual disturbance can be observed. For this case
the problem of finding an inequality like the Martens one has
not been completely solved however (cf. V. Dorofeev and J. de Graaf,
``Some maximality results for effect-valued measures,'' Indag.
Mathem. N.S. {\bf 8}, 349--369 (1997)). \\
\noindent
${}^{32}$D. Deutsch, ``Uncertainty in quantum measurements,''
Phys. Rev. Lett. {\bf 50}, 631-633 (1983); M. H. Partovi,
``Entropic formulation of uncertainty for quantum mechanics,''
Phys. Rev. Lett. {\bf 50}, 1883-1885 (1983); K. Kraus,
``Complementary observables and uncertainty relations,'' Phys.
Rev. {\bf D35}, 3070--3075 (1987).\\
\noindent
${}^{33}$H. Maassen and J. B. M. Uffink, ``Generalized entropic uncertainty
  relations,'' Phys. Rev. Lett. {\bf 60}, 1103--1106 (1988).\\
\noindent
${}^{34}$S. A. Werner and A. G. Klein, ``Neutron optics,'' in: K.
Sk\"old and D. L. Price eds, {\em Methods of Experimental Physics}
(Academic Press, Orlando, 1985), Vol. 23, Part A, pp. 259--337.\\
\noindent
${}^{35}$J. Summhammer, H. Rauch, and D. Tuppinger, ``Stochastic
and deterministic absorption in neutron-interferometric
experiments,'' Phys. Rev. {\bf A36}, 4447--4455 (1987).\\
\noindent
${}^{36}$A. Zeilinger, ``Complementarity in neutron
interferometry,'' Physica {\bf 137B}, 235--244 (1986).\\
\noindent
${}^{37}$R. Balian, {\em From microphysics to macrophysics}
(Springer-Verlag, Berlin, etc., 1991), Vol.~I, pp. 111--122.\\
\end{document}